\documentstyle[iopconf1]{article}

\begin{document}

\title{New Physics Near 1 TeV and Above}
\author{Roland E. Allen}

\affil{Department of Physics, Texas A\&M University \newline
College Station, Texas 77843}

\beginabstract
A new theory makes testable predictions: (1) Higgs fields have an
unconventional equation of motion. (2) Fermions have a second-order coupling
to gauge fields. (3) Fermion propagators are modified at high energy. (4)
There are new scalar bosons which are supersymmetric partners of spin 1/2
fermions. (5) Since W-bosons gravitate differently from fermions and
massless gauge bosons, there is a very small violation of the equivalence
principle. Other features of the theory have implications for cosmology,
including the values of cosmological parameters, a mechanism for
scale-invariant density fluctuations, and a candidate for dark matter.
\endabstract

\section{Introduction}

Superstring theory represents the best effort to extend conventional
physics, including Lorentz invariance, to the Planck scale~\cite{polchinski}.
Unfortunately, however, superstring theory has not yet yielded a single
prediction that can be tested against experiment.

Recently a different approach was introduced~\cite{superfield}, in which
Lorentz invariance is not presumed to hold at arbitrarily high energies.
Instead, a very simple form is postulated for the action, and Lorentz
invariance is then found to emerge as a good approximation at energies that
are well below the Planck scale.

This new theory turns out to lead to a large number of testable predictions,
several of which are presented here for the first time:

\begin{itemize}
\item  The Higgs boson has an unconventional equation of motion.

\item  Fermions have a second-order coupling to gauge fields.

\item  Fermion propagators are modified at high energy.

\item  There are new scalar bosons which are supersymmetric partners of spin
1/2 fermions.

\item  Since W-bosons gravitate differently from fermions and massless gauge
bosons, there is a very small violation of the equivalence principle.

\item  In cosmology, the density parameter $\Omega _{m}$ is less than one,
the universe is flat in the limit $t\rightarrow \infty $, and the
deceleration parameter approaches zero as $t\rightarrow \infty $.

\item  There is a mechanism for producing scale-invariant density
fluctuations which is different from the mechanism in standard inflationary
scenarios.

\item  The cold dark matter consists of superheavy, sterile, right-handed
neutrinos, produced by a nonthermal mechanism in the very early universe.
\end{itemize}

In the following sections, Ref. 2 will be cited as I, and equations from
this paper will be denoted with the prefix I -- e.g., I(3.1).

\section{Equation of Motion of Higgs Fields}

First consider an analogy. Suppose that an ordinary superfluid has an order
parameter
\begin{equation}
\psi _{s}=\exp \left( i\theta \right) n_{s}^{1/2}
\end{equation}
so that its velocity is $\overrightarrow{v_{s}}=m^{-1}\overrightarrow{\nabla
} \theta $. Suppose also that a nonrelativistic particle has a wavefunction
$\Psi _{p}\left( \overrightarrow{r}\right) =\exp \left( i \overrightarrow{p}
\cdot \overrightarrow{r}\right) \Psi _{p}\left( 0\right) $, so that its
velocity is $\overrightarrow{V_{p}}=m^{-1}\overrightarrow{p}$ (with the
masses taken to be equal for simplicity). Finally, suppose that an observer
is moving with the superfluid. In the frame of reference of this observer,
the particle has a velocity $\overrightarrow{v_{p}}= \overrightarrow{V_{p}}-
\overrightarrow{v_{s}}$ and a wavefunction $\psi _{p}$ given by
\begin{equation}
\Psi _{p}=\exp \left( i\theta \right) \psi _{p}  \label{8}
\end{equation}
where the prefactor $\exp \left( i\theta \right) $ represents a local
Galilean boost from $\overrightarrow{v_{p}}$ to $\overrightarrow{V_{p}}$.

In the present theory, $\psi _{s}$ is replaced by a GUT Higgs condensate
$\Psi _{s}$, and the particle wavefunction $\Psi _{p}$ is replaced by a
fundamental field $\Psi _{a}$. One of the principal new features of the
present theory is that $\Psi _{s}$ and $\Psi _{a}$ have two components
rather than one (in addition to the SO(10) gauge symmetry discussed below),
so they exhibit U(1)$\times $SU(2) rather than U(1) rotations in
4-dimensional spacetime. These rotations of the order parameter as a
function of $x^{\mu }$ are described by a 2$\times $2 matrix $U$:
\begin{equation}
\Psi _{s}=Un_{s}^{1/2}\eta
\end{equation}
where $\eta $ is a constant 2-dimensional vector. Since human observers live
in the GUT condensate $\Psi _{s}$, and are moving with it, one must
transform the initial field $\Psi _{a}$ to the field $\psi _{a}$ seen by
human observers. Just as in the analogy above, this is done by writing
\begin{equation}
\Psi _{a}=U\psi _{a}
\end{equation}
with the prefactor $U$ representing a boost in the U(1)$\times $SU(2)
rotations of $\psi _{a}$.

Let us begin with the bosonic excitations $\Phi _{b}$ defined by I(3.1),
written as
\begin{equation}
\Phi _{b}=U\phi _{b}\quad ,\quad \phi _{b}\left( x_{A},x_{B}\right)
=\sum_{r}\phi _{r}\left( x_{A}\right) \psi _{r}^{B}\left( x_{B}\right)
\end{equation}
in the notation of I(5.18). Repeating the arguments in Section 8 of I (with
appropriate modifications) leads to the Lagrangian density
\begin{equation}
{\cal L}_{b}=-h\left\{ h^{\mu \nu }\left[ \frac{1}{2m}\left( \widehat{D}
_{\mu }\phi \right) ^{\dagger }\left( \widehat{D}_{\nu }\phi \right) -\frac
{1}{2}mv_{\mu }^{\alpha }v_{\nu }^{\alpha }\phi ^{\dagger }\phi \right] +\frac
{1}{2}b^{\prime }\left( \phi ^{\dagger }\phi \right) ^{2}\right\} .
\end{equation}
(This is the Lorentzian version of I(8.32), but with $\Phi _{b}$ now written
in the form (5), and $b^{\prime }=\left( 2m\right) ^{-2}\overline{b}$. A
proper derivation of (6) will be given elsewhere.) Here $\phi $ is the
vector with components $\phi _{r}$ and
\begin{equation}
\widehat{D}_{\mu }=\partial _{\mu }+iA_{\mu }^{i}t_{i}+imv_{\mu }^{\alpha
}\sigma _{\alpha }\;.
\end{equation}
Also, $h^{\mu \nu }$ is the initial metric tensor of I(2.16), except that it
has been transformed from Euclidean to Lorentzian spacetime: $h^{\mu \nu
}=\eta ^{\mu \nu }$, where $\eta ^{\mu \nu }$ is the Minkowski metric tensor
$diag(-1,1,1,1)$.

In the present cosmological model [2-4], the ``superfluid velocity'' of the
GUT Higgs condensate has the form
\begin{equation}
v_{\alpha }^{\mu }=h^{\mu \nu }v_{\nu \alpha }=\lambda \delta _{\alpha
}^{\mu }
\end{equation}
in regions of spacetime where the gravitational field is weak. (Since
I(3.24) is not valid for a pure state, (8) provides the real justification
for the simple form of the Bernoulli equation I(3.26).) After an integration
by parts and the use of (8), the Lagrangian can then be simplified to
\begin{equation}
{\cal L}_{b}=-\frac{1}{2}h\,\left[ -\phi ^{\dagger }\left( \overline{m}
\,^{-1}g^{\mu \nu }D_{\mu }D_{\nu }+ie_{\alpha }^{\mu }\sigma ^{\alpha
}D_{\mu }\right) \phi +\frac{1}{2}b^{\prime }\left( \phi ^{\dagger }\phi
\right) ^{2}\right] +h.c.
\end{equation}
where $D_{\mu }=\partial _{\mu }+iA_{\mu }$, $A_{\mu }=A_{\mu }^{i}t_{i}$,
$\overline{m}=2\lambda ^{2}m$, $e_{\alpha }^{\mu }=v_{\alpha }^{\mu }$, and
``h.c.'' means ``Hermitian conjugate''. (Since I(8.31) is nonzero, this
equation replaces I(8.40).) As in I, $e_{\alpha }^{\mu }$ is interpreted as
the gravitational vierbein, so that $g^{\mu \nu }=\eta ^{\alpha \beta
}e_{\alpha }^{\mu }e_{\beta }^{\nu }=\lambda ^{2}h^{\mu \nu }$.

Also as in I, suppose that radiative corrections lead to small mass terms.
Let us focus on a particular field $\phi _{h}$ with a negative mass term:
\begin{eqnarray}
{\cal L}_{h}=\!\!\!\!\!\!  &-\frac{1}{2}g\left[ -\Phi _{h}^{\dagger }\left(
g^{\mu \nu }D_{\mu }D_{\nu }+i\overline{m}e_{\alpha }^{\mu }\sigma ^{\alpha
}D_{\mu }\right) \Phi _{h}-\mu _{h}^{2}\Phi _{h}^{\dagger }\Phi _{h}+\frac
{1}{2}\overline{b}\left( \Phi _{h}^{\dagger }\Phi _{h}\right) ^{2}\right] &
\nonumber \\
&+h.c.&
\end{eqnarray}
where $\Phi_{h}=\overline{m}\,^{-1/2}\lambda^{2}\phi_{h} $, $\overline{b}=
\left( 2m\right)
^{2}b^{\prime }$, and $A_{\mu }$ now represents just those gauge fields that
act on $\Phi _{h}$. We have used the fact that $g=\left( -\det g_{\mu \nu
}\right) ^{1/2}=\lambda ^{-4}\left( -\det h_{\mu \nu }\right) ^{1/2}=\lambda
^{-4}h$. For $\partial _{\mu }\Phi _{h}=0$, (10) becomes
\begin{equation}
{\cal L}_{h}= -g\left[ \Phi _{h}^{\dagger }\left(
g^{\mu \nu }A_{\mu }A_{\nu }+\overline{m}A_{\mu }e_{\alpha }^{\mu }\sigma
^{\alpha }\right) \Phi _{h}-\mu _{h}^{2}\Phi _{h}^{\dagger }\Phi _{h}+\frac
{1}{2} \overline{b}\left( \Phi _{h}^{\dagger }\Phi _{h}\right) ^{2}\right].
\end{equation}
Suppose that the total Lagrangian is postulated to contain both (11)
and a term having the form of its charge conjugate. Then the
contributions that are odd in $A_{\mu }$ will cancel in the condensed
phase, leaving
\begin{equation}
{\cal L}_{h}^{\prime }=-g\,\left[ \Phi _{h}^{\dagger }g^{\mu \nu }A_{\mu
}A_{\nu }\Phi _{h}-\mu _{h}^{2}\Phi _{h}^{\dagger }\Phi _{h}+\frac{1}{2}
\overline{b}\left( \Phi _{h}^{\dagger }\Phi _{h}\right) ^{2}\right] .
\end{equation}
In this way we arrive at a Lagrangian for the Higgs field $\Phi _{h}$ which has
the same form as in the standard electroweak theory.

Notice, however, that $g^{\mu \nu }$ is not to be regarded as a dynamical
variable, since it was merely used to rescale the field and volume element.
This means that there is no cosmological constant due to condensed Higgs
fields~\cite{superfield}.

Also, even though (12) has a conventional form, the equation of motion for a
Higgs field does not. Extremalizing ${\cal L}_{h}$ with respect to $\Phi
_{h}^{\dagger }$ gives
\begin{equation}
-\left( g^{\mu \nu }D_{\mu }D_{\nu }+i\overline{m}e_{\alpha }^{\mu }\sigma
^{\alpha }D_{\mu }\right) \Phi _{h}-\mu _{h}^{2}\Phi _{h}+\overline{b}\left(
\Phi _{h}^{\dagger }\Phi _{h}\right) \Phi _{h}=0.
\end{equation}
At very high energies and field strengths, this goes over to the usual form
\begin{equation}
\left[ -g^{\mu \nu }D_{\mu }D_{\nu }-\mu _{h}^{2}+\overline{b}\left( \Phi
_{h}^{\dagger }\Phi _{h}\right) \right] \Phi _{h}=0.
\end{equation}
At low energies and field strengths, however, it becomes
\begin{equation}
\left[ ie_{\alpha }^{\mu }\sigma ^{\alpha }D_{\mu }+\widetilde{\mu }-
\widetilde{b}\left( \Phi _{h}^{\dagger }\Phi _{h}\right) \right] \Phi _{h}=0
\end{equation}
where $\widetilde{\mu }=\mu _{h}^{2}/\overline{m}$ and $\widetilde{b}=
\overline{b}/\overline{m}$. According to the statement at the end of Ref. 3,
$\overline{m}$ must be greater than roughly 0.1 TeV. If $\mu _{h}$ is
$\sim $ 0.1 TeV, therefore, the prediction represented by (13) or (15) should be
testable in the near future.

Although (15) resembles the equation of motion of a fermion obtained in the
next section, $\Phi _{h}$ is a bosonic field which transforms as a scalar
under rotations.

The essential reason for the linear term in (13) is easily understood
through an analogy: Suppose that hypothetical observers in a two-dimensional
spacetime are aware that they are living in a superfluid (which corresponds
to the GUT Higgs condensate) and wish to determine whether it is rotating.
The conjectured rotation results from a vortex of strength $a$ centered on
$r=0$, so the superfluid velocity is given by I(7.5):
\begin{equation}
v_{s}=a/mr.
\end{equation}
Suppose also that these observers can measure the total energy of a
particle, with this energy known to have the form $E=\left[ \frac{1}{2}
m\left( v_{s}+v\right) ^{2}+V-\mu \right] $ if $\left( v_{s}+v\right) $ is
the velocity in a stationary frame of reference. Finally, recall that the
Bernoulli equation I(3.26) for an ordinary superfluid is $\frac{1}{2}
mv_{s}^{2}+V=\mu $ if the density $n_{s}$ is nearly constant. When the
energy is measured, it is found to be
\begin{equation}
E=\frac{1}{2}mv^{2}+mv_{s}v
\end{equation}
rather than simply $E=\frac{1}{2}mv^{2}$. The term that is linear in $v$ is
a signature that the observers are living in a rotating rather than a
stationary condensate.

In 4-dimensional spacetime, with the topology $R^{1}\times S^{3}$, the
vortex is replaced by its simplest generalization, an SU(2) instanton, and
the U(1) rotations of the order parameter in two dimensions are replaced by
SU(2) rotations. The two terms in the energy of (17) are completely
analogous to the first two terms in the Lagrangian (10) or the equation of
motion (13).

In short, the linear term in (13) is a signature of the SU(2) rotations of
the Higgs condensate $\Psi _{s}$ which are predicted by the present theory.

\section{Coupling of Fermions to Gauge Fields}

Because of the symmetry between bosons and fermions in the present theory,
(9) also holds for the fermion field $\psi $ of I(6.16). However, since
$\psi $ is an anticommuting Grassmann field with $\left( \psi ^{\dagger }\psi
\right) ^{2}=-\left( \psi ^{\dagger }\right) ^{2}\psi ^{2}=0$, the
self-interaction term vanishes, leaving
\begin{equation}
{\cal L}_{f}=-\frac{1}{2}h\,\left[ -\psi ^{\dagger }\left( \overline{m}
\,^{-1}g^{\mu \nu }D_{\mu }D_{\nu }+ie_{\alpha }^{\mu }\sigma ^{\alpha
}D_{\mu }\right) \psi \right] +h.c.
\end{equation}
or
\begin{equation}
{\cal L}_{f}=\frac{1}{2}g\,\psi ^{\prime \dagger }\left( ie_{\alpha }^{\mu
}\sigma ^{\alpha }D_{\mu }+\overline{m}\,^{-1}g^{\mu \nu }D_{\mu }D_{\nu
}\right) \psi ^{\prime }+h.c
\end{equation}
where $\psi ^{\prime }=\lambda ^{2}\psi $ and we have again used $g=\lambda
^{-4}h$. The equation of motion for massless fermions is then
\begin{equation}
\left[ ie_{\alpha }^{\mu }\sigma ^{\alpha }D_{\mu }+\overline{m}
\,^{-1}g^{\mu \nu }D_{\mu }D_{\nu }\right] \psi ^{\prime }=0.
\end{equation}

At low fermion energies and low field strengths, the second term can be
neglected, and we obtain the usual Dirac equation for massless fermions:
\begin{equation}
ie_{\alpha }^{\mu }\sigma ^{\alpha }D_{\mu }\psi ^{\prime }=0.
\end{equation}
On the other hand, at high field strengths there is a second-order coupling
to the gauge fields: If $\partial _{\mu }\psi ^{\prime }$ is still small
compared to $\overline{m}$, but $A_{\mu }$ is comparable to or larger than
$\overline{m}$, ${\cal L}_{f}$ contains a term
\begin{equation}
-g\,\psi ^{\prime \dagger }\left( \overline{m}\,^{-1}g^{\mu \nu
}A_{\mu }A_{\nu }\right) \psi ^{\prime }.
\end{equation}
This second-order coupling should be observable in both real and virtual
processes. For example, there will be seagull diagrams leading to the
production of two gauge bosons, just as for nonrelativistic electrons
interacting with photons. Notice, however, that (19) is still gauge
invariant, since both terms in ${\cal L}_{f}$ involve the covariant
derivative $D_{\mu }$.

\section{Fermion Propagators at High Energy}

The Lagrangian density for a free massless fermion is
\begin{equation}
{\cal L}_{f}=\frac{1}{2}g\,\psi ^{\prime \dagger }\left( ie_{\alpha }^{\mu
}\sigma ^{\alpha }\partial _{\mu }+\overline{m}\,^{-1}g^{\mu \nu }\partial
_{\mu }\partial _{\nu }\right) \psi ^{\prime }+h.c.
\end{equation}
For right- and left-handed fields coupled by a Dirac mass $m_{f}$, this
becomes
\begin{eqnarray}
{\cal L}_{D} &=&\psi _{R}^{\dagger }\left( i\sigma ^{\mu }\partial _{\mu }+
\overline{m}\,^{-1}\eta ^{\mu \nu }\partial _{\mu }\partial _{\nu }\right)
\psi _{R} \\
&&+\psi _{L}^{\dagger }\left( i\overline{\sigma }^{\mu }\partial _{\mu }+
\overline{m}\,^{-1}\eta ^{\mu \nu }\partial _{\mu }\partial _{\nu }\right)
\psi _{L} \\
&&-m_{f}\psi _{R}^{\dagger }\psi _{L}-m_{f}\psi _{L}^{\dagger }\psi _{R}
\end{eqnarray}
in a locally inertial coordinate system, with $\overline{\sigma }^{0}=\sigma
^{0}$ and $\overline{\sigma }^{k}=-\sigma ^{k}$. The terms which are
second-order in $\partial _{\mu }$ lead to a modified form for the fermion
propagator~\cite{CPT-98},
\begin{equation}
S_{0}=\left( -\overline{m}\,^{-1}\eta ^{\mu \nu }p_{\mu }p_{\nu }+\rlap{/}
p-m_{f}+i\epsilon \right)^{-1} .
\end{equation}
Again, this modification should be observable in both real and virtual
processes. For example, it will affect the values of running coupling
constants between the electroweak and Planck scales. With the supersymmetry
of the present theory included, it will be interesting to see whether
$\alpha _{1}^{-1}$, $\alpha _{2}^{-1}$, and $\alpha _{3}^{-1}$ meet at a
common energy near $10^{13}$ TeV.

\section{New Scalar Bosons}

The fundamental action I(2.7) of the present theory has an unconventional
kind of supersymmetry, with an equal number of scalar bosons and spin 1/2
fermions. (These are the only fundamental fields, with gauge bosons and
gravitons arising as excitations of the GUT Higgs condensate.) Some of the
initial bosonic fields undergo condensation, as exemplified by the minimal
scheme~[5-7]
\begin{equation}
SO(10)\stackrel{\bf 16}{\rightarrow }SU(5)\stackrel{\bf 45}{\rightarrow }
SU(3)\times SU(2)\times U(1)\stackrel{\bf 10}{\rightarrow }SU(3)\times U(1)
\end{equation}
with the required multiplets of bosonic fields indicated for each stage of
symmetry-breaking. The corresponding energy scales are presumably in the
vicinity of $m_{P}\sim 10^{15}$ TeV to $m_{GUT}\sim 10^{13}$ TeV for the
first two stages, and $m_{ew}\sim 0.25$ TeV for the last. The formation of
the condensate $\Psi _{s}$ begins with the first stage, but at low energies
all three sets of condensed bosonic fields contribute.

Suppose that there are initially 3 families of fermions, each with 16 fields
(corresponding to the spinorial representation of SO(10)). There are then
the same number of fundamental bosonic fields (each with two complex
components), for a total of $3\times 16\times 2\times 2=192$ real degrees of
freedom. Many of these will be lost near the GUT scale, due to condensation
and the growth of gauge-boson masses. Four of them will participate in the
symmetry-breaking of the standard model at the electroweak scale. Others,
however, will behave as normal scalar bosons.

If the supersymmetry of the present theory is to prevent the usual quadratic
divergence for the electroweak Higgs boson, then there must be new scalar
bosons with masses near 1 TeV. The present theory thus predicts
supersymmetric partners of spin 1/2 fermions which may be observable at
Tevatron or LHC energies.

\section{W-bosons and the Equivalence Principle}

The first term in (12) leads, as usual, to a mass term in the equation of
motion of $W^{\pm }$ and $Z^{0}$ bosons. Recall, however, that the original
form is
\begin{equation}
{\cal L}_{w}=-\lambda ^{-2}h\,\Phi _{h}^{\dagger }h^{\mu \nu }A_{\mu
}A_{\nu }\Phi _{h}.
\end{equation}
It is valid to assume that $\lambda ^{-2}h\,h^{\mu \nu }\approx g\,g^{\mu
\nu }$ only in those regions of spacetime where the gravitational field is
weak. (The assumption $\,g^{\mu \nu }=\lambda ^{2}h^{\mu \nu }$ \cite{CPT-98}
is supposed to hold only in such regions -- e.g., in interstellar space
during the current epoch of cosmological time.) In a gravitational field,
$g^{\mu \nu }$ is not proportional to $h^{\mu \nu }$, and ${\cal L}_{w}$ will
consequently not assume its conventional form in a locally inertial
coordinate system.

Since ordinary matter contains only a very small concentration of virtual
W-bosons, the resulting violation of the equivalence principle should also
be very small, but it is in principle observable.

\section{Cosmological Parameters}

In discussing the implications for cosmology, let us begin with a
hypothetical universe which contains no matter or radiation. Since it costs
action (in a Euclidean picture) or energy (in a Lorentzian picture) to form
new instantons, the topological charge is fixed as a function of the time
$x^{0}$. According to Ref. 4, this implies that
\begin{equation}
R(t)=\left( m/a\right) \left( x^{0}\right) ^{2}\quad \mbox{with}\quad
dt=e_{0}dx^{0}
\end{equation}
where
\begin{equation}
e_{0}=e_{\mu =0}^{\alpha =0}\quad \mbox{and}\quad e^{0}=e_{\alpha =0}^{\mu
=0}\quad \mbox{so that }\quad e_{0}e^{0}=1.
\end{equation}
The arguments of Refs. 3 and 4 also imply that $e_{0}=\left( m/a\right)
x^{0} $ for a universe in which the GUT Higgs condensate $\Psi _{s}$ is
fully formed. Since the density of matter $\rho $ declines as $R^{-3}$, both
of the above assumptions are valid in the remote future, and we have
\begin{equation}
dR/dt=2\quad \mbox{in the limit}\quad t\rightarrow \infty .
\end{equation}

The present theory thus predicts a coasting universe with
\begin{equation}
q\equiv -\frac{d^{2}R/dt^{2}}{H^{2}R}\rightarrow 0\quad
\end{equation}
as $t\rightarrow \infty $. In the early universe, however, gravity dominates
and Einstein's field equations are satisfied to a very good approximation.
Suppose that $\Omega _{m}\approx 0.3$~\cite{freedman,netabahcall}, where
\begin{equation}
\Omega _{m}\equiv \frac{\rho }{\rho _{c}}\quad ,\quad \rho _{c}\equiv \left(
\frac{3}{8\pi G}\right) H^{2}\quad ,\quad \mbox{and}\quad H\equiv \frac{dR/dt}
{R}.
\end{equation}
Suppose also that $h\approx 0.65$~\cite{freedman,netabahcall}
where $H_{0}=100\,h$ km s$^{-1}$Mpc$
^{-1}$. An interpolative model then gives~\cite{mondragon}
\begin{equation}
q_{0}\approx 0.15\quad \mbox{and}\quad t_{0}\approx 12.5\;\mbox{Gyr}.
\end{equation}
One could, of course, add a cosmological constant which would make $q_{0}$
negative and would yield a larger age for the universe, but such a term
would be just as {\it ad hoc} in the present theory as it is in standard
cosmology. Recent observations of Ia supernovae suggest an accelerating
universe~\cite{perlmutter,garnavich}, or a negative value of $q_{0}$, but
there are still uncertainties in the interpretation of this data [13-15].

\section{Scale-Invariant Density Fluctuations}

In the very early universe, there is again negligible matter and radiation,
so the form (30) still holds. As described in Refs. 2 and 4, this form
follows from the assumption of a cosmological instanton centered (in
four-dimensional Euclidean spacetime) on the point $x^{0}=0$. The
``superfluid velocity'' $v_{a}^{k}$ (with $k,a=1,2,3$) diverges at this
point, with
\begin{equation}
v_{a}^{k}=\delta _{a}^{k}\,\overline{a}/mx^{0}.
\end{equation}
However, this singularity is precisely analogous to the singularity at the
center of a vortex in an ordinary superfluid, with
\begin{equation}
n_{s}=\Psi _{s}^{\dagger }\Psi _{s}\rightarrow 0\quad \mbox{as}\quad
t\rightarrow 0.
\end{equation}
The Big Bang singularity at $t=0$ is thus physically admissible in the
present theory.

Since the initial cosmological instanton has a large strength $\overline{a}$,
it at first appears that the condensate density $n_{s}$ will remain very
small until a late time $x^{0}$ $\sim \overline{a}/m$: With $\Psi
_{s}=n_{s}^{1/2}U\eta$, $r=x^{0}$, and the scalings $\rho =r/\xi $,
$f=\left( n_{s}/\overline{n}_{s}\right) ^{1/2}$, $\xi =\left( 2m\mu \right)
^{-1/2}$, $\overline{n}_{s}=\mu /b$, the generalized Bernoulli equation
I(3.26) becomes
\begin{equation}
-\frac{1}{f}\frac{1}{\rho ^{3}}\frac{d}{d\rho }\left( \rho ^{3}\frac{df}
{d\rho }\right) +\frac{3\overline{a}^{2}}{\rho ^{2}}+f^{2}=1
\end{equation}
in the original Euclidean picture of I. The asymptotic solution is
\begin{equation}
\quad \quad f=C\rho ^{n}\quad \mbox{as}\quad \rho \rightarrow 0
\end{equation}
where $C$ is a constant and $n=\left( 1+3\overline{a}^{2}\right) ^{1/2}-1$,
so
\begin{equation}
n_{s}\propto \left( x^{0}\right) ^{2n}\quad \mbox{as}\quad x^{0}\rightarrow
0.
\end{equation}
For $x^{0}$ $\ll \overline{a}/m$, therefore, $n_{s}$ is very flat and very
nearly equal to zero.

The above behavior, however, assumes no topological defects other than the
instanton and no fields other than $\Psi _{s}$. As mentioned in I, there may
also be monopole-like defects which act as sources of the U(1) velocity $
v_{0}^{0}=e^{0}$. Although it costs action or energy to form such defects,
there is a net saving because the term involving $v_{\alpha }^{0}$ tends to
cancel the term involving $v_{\alpha }^{k}$ in the Bernoulli equation
\begin{equation}
P+\frac{1}{2}m\left( -v_{\alpha }^{0}v_{\alpha }^{0}+v_{\alpha
}^{k}v_{\alpha }^{k}\right) +V=\mu
\end{equation}
so the GUT\ Higgs field can condense far more rapidly. (The above equation
is the Lorentzian form of I(3.26), with (38) as the Euclidean version when
$v_{\alpha }^{0}=0$.) Suppose that, very near $x^{0}=0$, such defects are
produced by quantum fluctuations at a rapid rate which is proportional to
the 3-volume and constant with respect to $x^{0}$. Then $e^{0}$ grows
linearly with $x^{0}$:
\begin{equation}
e^{0}=\frac{1}{2}\overline{H}\,x^{0}\quad \mbox{with}\quad \quad \overline{H}
=\mbox{constant}.
\end{equation}
This means that $e_{0}=2/\left( \overline{H}x^{0}\right) $, so that
$dt=e_{0}dx^{0}=2\,dx^{0}/\left( \overline{H}x^{0}\right) $ and
\begin{equation}
t=\left( 2/\overline{H}\right) \log \,x^{0}+\mbox{constant}.
\end{equation}
It follows that $x^{0}\left( t\right) =x^{0}\left( 0\right) \exp \left(
\overline{H}t/2\right) $ and, according to (30),
\begin{equation}
R(t)=R(0)\exp \left( \overline{H}t\right) .
\end{equation}
There is then an exponential increase in the cosmic scale factor $R$ as a
function of the proper time $t$, just as in inflationary scenarios. Since
the fermion fields $\psi $, the scalar boson fields $\phi $, the gauge
fields $A_{\mu }^{i}$, and the gravitational field $g_{\mu \nu }$ experience
$t$ as the physical time (to a good approximation), (44) provides a potential
mechanism for producing
the scale-invariant Harrison-Zel'dovich density fluctuations which are
observed in the cosmic background radiation~[16-19].

\section{Superheavy Sterile Neutrinos as Dark Matter}

Since the present theory contains an SO(10) grand unified gauge theory, it
predicts an additional right-handed field $\nu _{R}$ for each family of
fermions, in addition to the 15 fields of the standard model. Suppose that
this additional field has a large Majorana mass $M_{\nu }$ which results
from a Yukawa coupling to a Higgs field $H$ near the GUT scale, involving a
term with the form

\begin{equation}
\nu _{R}C\nu _{R}H.
\end{equation}
Such a term may arise from radiative corrections, and it will lead, through
the seesaw mechanism, to naturally small masses for ordinary left-handed
neutrinos [5-7].

In addition, however, there will be superheavy right-handed neutrinos.
Sterile neutrinos of this kind will interact only gravitationally, so
they will be stable for cosmologically long times. Such neutrinos are a
candidate for dark matter, provided they can be produced in the right
abundance in the early universe~\cite{nanopoulos,fuller}.

In the ``inflationary'' scenario of the preceding section, $\langle H\rangle
$ is very small near $t=0$. It follows that the right-handed neutrinos $\nu
_{R}$ will have very small masses during the
period when $R(t)\propto \exp \left( \overline{H}t\right) $, and that they
can be easily produced by quantum fluctuations. After this period, formation
of the condensate $\langle H\rangle $ causes each $\nu _{R}$ to grow a mass
near the GUT scale, allowing these particles to play the role of cold dark
matter in subsequent structure formation.

Although superheavy sterile neutrinos do not appear to be directly observable
in terrestrial dark-matter or neutrino-oscillation experiments, they might have a noticeable
effect on proton decay.

\section{Other Features of the Theory}

The present theory begins with the simplest imaginable action for a world
which contains both bosons and fermions. The continuum version is given by
I(2.7):
\begin{equation}
S=\int d^{D}x\left[ \frac{1}{2m}\partial ^{M}\Psi ^{\dagger }\partial
_{M}\Psi -\mu \Psi ^{\dagger }\Psi +\frac{1}{2}b\left( \Psi ^{\dagger }\Psi
\right) ^{2}\right] .
\end{equation}
$\Psi $ is called a statistical superfield because its Euclidean action has
the same form as $\beta H$ for standard models in statistical mechanics. The
Euclidean path integral, of course, has the same form as a partition
function in statistical mechanics:
\begin{equation}
Z~=~\int {\cal D}\Psi \; {\cal D}\Psi ^{\dagger } \, e^{-S}.
\end{equation}

Although the initial action is extremely simple, it can lead to the full
richness of nature because of spontaneous symmetry-breaking and the
formation of topological defects. Three types of topological defects are of
central importance: the cosmological U(1)$\times $SU(2) instanton discussed
in Refs. 2 and 4; an internal instanton which gives rise to the SO(10) gauge
symmetry; and Planck-scale instantons which account for the curvature of
both gauge fields and the gravitational field.

\bigskip In addition to its simplicity, the present theory has several other
favorable features, including the following:

\begin{itemize}
\item  In conventional higher-dimensional theories, it is difficult to
understand why the internal space with $\left( D-4\right) $ dimensions is a
compact manifold with a size comparable to the Planck length $\ell _{P}$. In
the present theory, however, the effective internal space is automatically a
d-sphere with volume
\begin{equation}
V_{d}\sim \ell _{P}^{d}
\end{equation}
as indicated in I(7.23).

\item  With $d=9$, one automatically obtains an SO(10) gauge group. The
present theory thus implies neutrino masses, with about the right magnitudes
to explain recent and established experiments involving atmospheric and
solar neutrinos [22-27].

\item  Family replication results naturally if the initial group $G$ of I is
slightly larger than Spin(10). I.e., the gauge group is determined by the
nature of the internal space $V_{d}$, but the number of fermion species is
determined separately, by the number of components in the original fermion
field $\Psi _{f}$ of I(2.13).

\item  The theory is economical, in that it does not lead to vast numbers of
new fields or particles. At the same time, it implies a wide variety of new
phenomena in high energy physics and astrophysics.
\end{itemize}

\section{Conclusion}

The present theory leads to a large number of testable predictions. Perhaps
the most promising for the near future is represented by (13): a scalar
Higgs boson with an unconventional equation of motion. In addition, the
supersymmetry of the theory implies a set of new scalar bosons near and
above 1 TeV. The other predictions listed in Section 1 are also directly
relevant to experiments and observations planned for the next 5-10 years,
which promise to be an exceptionally productive period for both
astrophysics and high energy physics.

\section*{Acknowledgements}

This work was supported by the Robert A. Welch Foundation. I thank R.C. Webb
for the suggestion in the last paragraph of Section 9.


\begin{thebibliography}{99}
\bibitem{superfield}  Allen R E 1997 {\it Int. J. Mod. Phys. A} 12, 2385;
hep-th/9612041. This paper will be cited as I, and equations from this paper
will be denoted with the prefix I -- e.g., I(5.18).
\end{thebibliography}
\end{document}